\documentclass[aps,prl,amsmath,amssymb,footinbib,showpacs,twocolumn,superscriptaddress]{revtex4-1}
\usepackage{amsmath}
\usepackage{amssymb}
\usepackage{amsthm}
\usepackage{setspace}
\usepackage{graphicx}
\usepackage{braket}
\usepackage{mathrsfs}
\usepackage{float}
\usepackage[colorlinks = true,linkcolor = red,urlcolor  = blue,citecolor = blue,anchorcolor = blue]{hyperref}
\usepackage[utf8]{inputenc}
\usepackage[english]{babel}
\usepackage{bm}

\begin{document}

\title{The sign of longitudinal magnetoconductivity and the planar Hall effect in Weyl semimetals}

\author{Girish Sharma}
\affiliation{School of Basic Sciences, Indian Institute of Technology Mandi, Mandi-175005, India}

\author{S. Nandy}
\affiliation{Department of Physics, University of Virginia, Charlottesville, VA 22904, USA}

\author{Sumanta Tewari}
\affiliation{Department of Physics and Astronomy, Clemson University, Clemson, South Carolina 29634, USA}

\begin{abstract}
The manifestation of chiral anomaly in Weyl semimetals typically relies on the observation of longitudinal magnetoconductance (LMC) along with the planar Hall effect, with a specific magnetic field and angle dependence. Here we solve the Boltzmann equation in the semiclassical regime for a prototype of a Weyl semimetal, allowing for both intravalley and intervalley scattering, along with including effects from the orbital magnetic moment (OMM), in a geometry where the electric and magnetic fields are not necessarily parallel to each other. We construct the phase diagram in the relevant parameter space that describes the shift from positive to negative LMC in the presence of OMM and sufficiently strong intervalley scattering, as has been recently pointed out for only parallel electric and magnetic fields. On the other hand, we find that the chiral anomaly contribution to the planar Hall effect always remains positive (unlike the LMC) irrespective of the inclusion or exclusion of OMM, or the strength of the intervalley scattering. Our predictions can be directly tested in experiments, and may be employed as new diagnostic procedures to verify chiral anomaly in Weyl systems.
\end{abstract}

\maketitle

\section{Introduction}
The Weyl equation, once only constrained to the realm of high-energy physics~\cite{peskin1995introduction}, has now  found relevance in describing the low-energy quasiparticle excitations of massless three-dimensional chiral electron fluids~\cite{murakami2007phase,murakami2007tuning,volovik2003universe,burkov2011topological,burkov2011weyl,wan2011topological,xu2011chern,yang2011quantum}. The Weyl equation specifically describes the low-energy excitation in the vicinity of doubly degenerate band touching point. Two such points located at isolated positions in the momentum
space defines a minimal model of the simplest possible Weyl semimetal (WSM). Topological properties of Weyl semimetals are encapsulated in the fact that these diabolic points act as the source and sink of Abelian Berry curvature, and are protected by a non-trivial integral Chern number $\mathcal{C}=\pm 1$~\cite{nielsen1981no,nielsen1983adler,xiao2010berry}. Higher Chern numbers are also possible in multi-Weyl semimetals with non-linear dispersion~\cite{fang2012multi}. It also follows that all Weyl semimetals must break either time-reversal (TR) symmetry or spatial-inversion (SI) symmetry, in order for the Berry flux to have a non-trivial distribution inside the Brillouin zone~\cite{xiao2010berry}. Many experimental probes (such as the anomalous Hall effect)~\cite{yang2011quantum,burkov2014anomalous} are devoted to the measurement of the Chern number. 

Anomalous Hall effect has not been uncommon in condensed matter even prior to the discovery of WSMs (see Ref.~\cite{nagaosa2010anomalous} and references therein). However, WSMs offer more interesting physics, unlike previous condensed matter platforms, due to the realization of quantum anomalies, the most prominent among them being the  chiral or Adler-Bell-Jackiw anomaly~\cite{adler1969axial,bell1969pcac,aji2012adler,zyuzin2012weyl,zyuzin2012weyl,son2012berry,goswami2013axionic,goswami2015optical, fukushima2008chiral}. Weyl fermions always appear in pairs (also termed as left/right flavored), and in the absence of any external gauge or gravitational field coupling, the numbers of left-handed and right-handed Weyl fermions are separately conserved. However, in the presence of background gauge fields, such as an electromagnetic field, the separate number conservation laws no longer holds true, which is a result of chiral anomaly. The current conservation law dictates $\partial_\mu j^\chi_\mu = \frac{e^2}{h^2} \mathbf{E}\cdot \mathbf{B}$, indicating charge pumping from one Weyl node to the other as long as $\mathbf{E}\cdot \mathbf{B}\neq 0$. Chiral anomaly has become one of the most prominent effects with origin in high energy physics that has been verified in condensed matter physics experiments, with possible experimental signatures expected to arise from positive longitudinal magnetoconductance (LMC)~\cite{son2013chiral,kim2014boltzmann,zyuzin2017magnetotransport,he2014quantum,liang2015ultrahigh,zhang2016signatures,li2016chiral,xiong2015evidence,hirschberger2016chiral}, planar Hall effect~\cite{nandy2017chiral,kumar2018planar,yang2019w,li2018giant,chen2018planar,li2018giant2,yang2019frustration,pavlosiuk2019negative,singha2018planar}, optical gyrotropy~\cite{goswami2015optical}, and thermopower~\cite{lundgren2014thermoelectric,sharma2016nernst,das2019berry}.

Recently it has been realized that positive longitudinal magnetoconductance is neither a necessary, nor a sufficient condition to prove the existence of chiral anomaly in Weyl semimetals. Specifically, positive LMC can arise even in the absence of any Weyl nodes (such as observed in ultraclean PdCoO$_2$)~\cite{kim2009fermi,noh2009anisotropic,kikugawa2016interplanar}. Secondly, experimentally it is known that jetting effect can result in false positive LMC~\cite{dos2016search} due to extrinsic reasons even in the absence of chiral anomaly, although in a recent work~\cite{liang2018experimental}, it has been shown that by careful measurement of voltage drops along the mid ridge and edges of the sample, one can eliminate this effect. On the  theoretical front, it is also now established that Weyl nodes may not always result in a positive LMC. For strong magnetic fields (when Landau quantization is relevant), LMC can be either positive or negative for short-range scatterers, while it is usually positive for charged impurities~\cite{goswami2015axial,lu2015high,chen2016positive,zhang2016linear,shao2019magneto,li2016weyl,ji2018effect}. In the weak magnetic field regime, it was believed that the LMC is  positive~\cite{spivak2016magnetotransport,das2019linear,imran2018berry,dantas2018magnetotransport,johansson2019chiral,grushin2016inhomogeneous,cortijo2016linear,sharma2017chiral}, however a recent study by Knoll \textit{et al.} (Ref.~\cite{knoll2020negative}) has now shown that unlike the earlier claims, the LMC can be negative for sufficiently strong intervalley scattering when orbital magnetic moment (OMM) effects are included.

\begin{figure}
	\includegraphics[width=0.5\columnwidth]{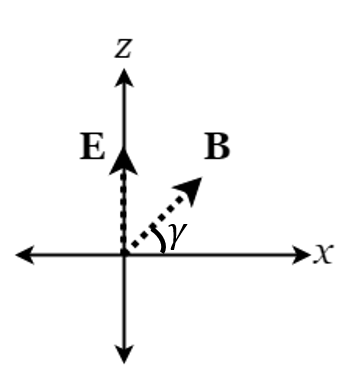}
	\caption{Schematic diagram of the  geometry followed in this work. The electric field points along the $\hat{z}-$direction, while the magnetic field is rotated along the $xz-$plane. $\gamma$ is the angle the magnetic field with respect to the $\hat{x}-$axis. }
	\label{Fig_schematic}
\end{figure}

While Ref.~\cite{knoll2020negative} correctly solves the semiclassical Boltzmann equation  in the specific geometry when the electric ($\mathbf{E}$) and magnetic ($\mathbf{B}$) fields are parallel to each other, it still remains to be understood as to how are the conclusions modified when these fields are not necessarily parallel to each other (the actual and more general requirement for chiral anomaly). This is an important question to be addressed, pertinent to several recent and upcoming experiments on Weyl semimetals. Secondly, the consequence of intervalley scattering in the presence of OMM on the planar Hall effect has not yet been explored. This is another equally important problem, given the fact that several works have reported the measurement of planar Hall effect on Weyl systems~\cite{kumar2018planar,yang2019w,li2018giant,chen2018planar,li2018giant2,yang2019frustration,pavlosiuk2019negative,singha2018planar}. The current manuscript addresses these two questions, and solves the Boltzmann equation for a prototype of a Weyl semimetal in setup where the $\mathbf{E}$ and $\mathbf{B}$ fields are non-collinear. Without loss of generality, in our work we will fix $\mathbf{E}$ pointing along the $\hat{z}-$axis, while the $\mathbf{B}$ rotates in the $xz-$plane (see Fig.~\ref{Fig_schematic}). 
Note that from elementary field-theory calculations~\cite{fukushima2008chiral}, the chiral chemical potential ($\mu_5$) created by the external $\mathbf{E}$ and $\mathbf{B}$ fields in the presence of intervalley scattering is $\mu_5 = 3v_F^3 e^2 \tau_i \mathbf{E}\cdot \mathbf{B}/4\hbar^2 \mu^2$, where $v_F$, $\tau_i$, and $\mu$ denote the Fermi velocity, scattering time, and the chemical potential, respectively. The corresponding current is given by $\mathbf{j} = e^2 \mu_5 \mathbf{B}/ 2\pi^2$, which immediately gives us LMC as well as the planar Hall effect. Although this simple analysis predicts the angular dependence of the longitudinal magnetoconductance as well as the planar Hall conductance, it is by no means obvious as to how the LMC and planar Hall conductance will behave as a function of magnetic field in a geometry when the fields are non-collinear (whether the chiral anomaly contribution is positive or negative). A detailed calculation in the general setup including intravalley scattering, intervalley scattering, and the OMM effects in the semiclassical regime  of experimental interest, is therefore imperative. 
In fact, as we shall show, many interesting and non-trivial features  emerge from our analysis. 

Specifically, we find that as the angle between the electric and magnetic field varies from $\pi/2$ to $0$, the LMC changes from negative to positive (at a particular angle depending on the relative intervalley scattering $\alpha_i$) when (i) OMM is present, and (ii) $\alpha_i$ is lesser than the angle dependent critical value $\alpha_{ic}^\gamma$. Specifically, we also trace out the phase diagram in $\gamma-\alpha_i$ parameter space that describes the shift from positive to negative LMC ($\gamma$ defined as the angle the ${B}-$field makes with the $x$-axis while the electric field is along the $z$-axis, see Fig.~\ref{Fig_schematic}). When $\alpha_i>\alpha_{ic}^\gamma$, we find that the LMC is always negative.
For completeness we also perform all the calculations without the inclusion of OMM, where we always find a positive LMC at all angles of the $B-$field, as well as at all values of the intervalley strength $\alpha_i$. Interestingly, the planar Hall conductance always increases and remains positive with the magnetic field (unlike the LMC) irrespective of the presence or absence of OMM or the relative strength of the intervalley scattering.  Our predictions can be directly tested in experiments, and may be employed as new diagnostic procedures to verify chiral anomaly in Weyl systems.

Before closing this section, we also briefly comment on the convention of LMC and the planar Hall conductance followed in this work. Typically in WSMs, the LMC, $\sigma_{zz}(B) = \sigma_{zz}^0 + \delta \sigma_{zz}(B)$, where $\sigma_{zz}^0$ and $\delta \sigma_{zz}(B)$ refer to the zero-field value and the chiral anomaly contribution respectively. If the deviation $\delta \sigma_{zz}(B)/\sigma_{zz}^0$ increases (decreases) as a function of magnetic field, we term it as positive (negative) LMC. An equivalent way (followed here) is to relate $\sigma_{zz}(B)/\sigma_{zz}^0 >1$ (or $<1$) to positive (negative) LMC. For the planar Hall conductance, this issue does not arise as $\sigma_{xz}(B=0)=0$, and thus the net planar Hall conductivity comes solely from the anomaly contribution. 

\section{Theoretical modelling}
We will focus on a simple prototype model of a WSM, consisting of two diabolic Weyl points  of opposite chiralities ($\chi=\pm 1$), ignoring the non-universal corrections due to band curvature far away from the nodes. The Hamiltonian expanded around each Weyl point can be expressed as 
\begin{align}
H = \chi \hbar v_F \mathbf{k}\cdot\boldsymbol{\sigma},
\label{Eq_Hweyl}
\end{align}
where $v_F$ is the Fermi velocity, $\mathbf{k}$ is the momentum measured relative to the Weyl point, and $\boldsymbol{\sigma}$ is the vector of the Pauli matrices. 

We will resort to the Boltzmann theory to study transport in the presence of weak electric and magnetic fields, and thus the Landau quantization regime will not be relevant for our discussion. A phenomenological Boltzmann equation for the non-equilibrium distribution function $f^\chi_\mathbf{k}$ can be written as 
\begin{align}
\left(\frac{\partial}{\partial t} + \dot{\mathbf{r}}^\chi\cdot \nabla_\mathbf{r}+\dot{\mathbf{k}}^\chi\cdot \nabla_\mathbf{k}\right)f^\chi_\mathbf{k} = \mathcal{I}_{{col}}[f^\chi_\mathbf{k}],
\label{Eq_boltz1}
\end{align}
where the collision term on the right-hand side incorporates the effect of impurity scattering.
The presence of Berry flux modifies the semiclassical dynamics of Bloch electrons in the presence of electric ($\mathbf{E}$) and magnetic ($\mathbf{B}$) fields as~\cite{son2012berry} 
\begin{align}
\dot{\mathbf{r}}^\chi &= \mathcal{D}^\chi \left( \frac{e}{\hbar}(\mathbf{E}\times \boldsymbol{\Omega}^\chi + \frac{e}{\hbar}(\mathbf{v}^\chi\cdot \boldsymbol{\Omega}^\chi) \mathbf{B} + \mathbf{v}_\mathbf{k}^\chi)\right) \nonumber\\
\dot{\mathbf{p}}^\chi &= -e \mathcal{D}^\chi \left( \mathbf{E} + \mathbf{v}_\mathbf{k}^\chi \times \mathbf{B} + \frac{e}{\hbar} (\mathbf{E}\cdot\mathbf{B}) \boldsymbol{\Omega}^\chi \right),
\end{align}
where we reserve the index $\chi$ indicating that the particular quantity pertains to a specific chirality. The quantity $\mathbf{v}_\mathbf{k}^\chi$ is the band velocity, $\boldsymbol{\Omega}^\chi = -\chi \mathbf{k} /2k^3$ is the Berry curvature, and $\mathcal{D}^\chi = (1+e\mathbf{B}\cdot\boldsymbol{\Omega}^\chi/\hbar)^{-1}$ is the factor by which the phase space volume is modified. Further, the self-rotating Bloch wavepacket also gives rise to an orbital magnetic moment (OMM), given by $\mathbf{m}^\chi_\mathbf{k} = -e v_F \mathbf{k}/2k^2$. In the presence of magnetic field, the OMM shifts the energy dispersion as $\epsilon^{\chi}_{\mathbf{k}}\rightarrow \epsilon^{\chi}_{\mathbf{k}} - \mathbf{m}^\chi_\mathbf{k}\cdot \mathbf{B}$. Particular importance must be paid to the signs in the definition of the Berry curvature and OMM, which are crucial in obtaining correct results. 

The collision integral $\mathcal{I}_{{col}}[f^\chi_\mathbf{k}]$ can be expressed as 
\begin{align}
\mathcal{I}_{{col}}[f^\chi_\mathbf{k}] = \sum\limits_{\chi'}\sum\limits_{\mathbf{k}'} W^{\chi\chi'}_{\mathbf{k},\mathbf{k}'} (f^{\chi'}_{\mathbf{k}'} - f^\chi_\mathbf{k}),
\end{align}
where the scattering rate $W^{\chi\chi'}_{\mathbf{k},\mathbf{k}'}$ in the first Born approximation is given by~\cite{bruus2004many} 
\begin{align}
W^{\chi\chi'}_{\mathbf{k},\mathbf{k}'} = \frac{2\pi}{\hbar} \frac{n}{\mathcal{V}} |\langle \psi^{\chi'}_{\mathbf{k}'}|U^{\chi\chi'}_{\mathbf{k}\mathbf{k}'}|\psi^\chi_\mathbf{k}\rangle|^2 \delta(\epsilon^{\chi'}_{\mathbf{k}'}-\epsilon_F)
\label{Eq_W_1}
\end{align}
In the above expression $n$ is the impurity concentration, $\mathcal{V}$ is the system volume, $|\psi^\chi_\mathbf{k}\rangle$ is the Weyl spinor wavefunction (obtained from diagonalizing Eq.~\ref{Eq_Hweyl}), $U^{\chi\chi'}_{\mathbf{k}\mathbf{k}'}$ is the scattering potential profile, and $\epsilon_F$ is the Fermi energy. The exact nature of $U^{\chi\chi'}_{\mathbf{k}\mathbf{k}'}$ is fixed by the nature of impurities. For this work, we restrict our attention to non-magnetic point-like scatterers, but make a clear distinction  between intervalley and intravalley scattering. Therefore the scattering matrix becomes momentum-independent, involves no off-diagonal entries in the spinor space, but retains a chirality dependence, i.e.,  $U^{\chi\chi'}_{\mathbf{k}\mathbf{k}'} = U^{\chi\chi'}\mathbb{I}$.

The distribution function is assumed to take the form $f^\chi_\mathbf{k} = f_0^\chi + g^\chi_\mathbf{k}$, where $f_0^\chi$ is the equilibrium Fermi-Dirac distribution function and $g^\chi_\mathbf{k}$ indicates the deviation from equilibrium. 
In the steady state, the Boltzmann equation (Eq.~\ref{Eq_boltz1}) takes the following form 
\begin{align}
-e \mathcal{D}^\chi&\left[\left(\frac{\partial f_0^\chi}{\partial \epsilon^\chi_\mathbf{k}}\right) \mathbf{E}\cdot \left(\mathbf{v}^\chi_\mathbf{k} + \frac{e\mathbf{B}}{\hbar} (\boldsymbol{\Omega}^\chi\cdot \mathbf{v}^\chi_\mathbf{k}) \right)+ \frac{\mathbf{v}^\chi_\mathbf{k} \times \mathbf{B}}{\hbar} \cdot \nabla_\mathbf{k} g^\chi_{\mathbf{k}}\right]\nonumber\\
 &= \sum\limits_{\chi'}\sum\limits_{\mathbf{k}'} W^{\chi\chi'}_{\mathbf{k}\mathbf{k}'} (g^\chi_{\mathbf{k}'} - g^\chi_\mathbf{k})
 \label{Eq_boltz2}
\end{align}
The deviation $g^\chi_\mathbf{k}$ is assumed to be proportional to the applied electric field 
\begin{align}
g^\chi_\mathbf{k} = e \left(-\frac{\partial f_0^\chi}{\partial \epsilon^\chi_\mathbf{k}}\right) \mathbf{E}\cdot \boldsymbol{\Lambda}^\chi_\mathbf{k}
\end{align}
We will fix the direction of the applied external electric field to be along $+\hat{z}$, i.e., $\mathbf{E} = E\hat{z}$. Therefore only ${\Lambda}^{\chi z}_\mathbf{k}\equiv {\Lambda}^{\chi}_\mathbf{k}$, will be relevant.
Further, we rotate the magnetic field along the $xz$-plane such that it makes an angle $\gamma$ with respect to the $\hat{x}-$axis, i.e., $\mathbf{B} = B(\cos\gamma,0,\sin\gamma)$. Therefore, when $\gamma=\pi/2$, the electric and magnetic fields are parallel to each other (the typical longitudinal magnetoconductance geometry). When $\gamma\neq \pi/2$, the electric and magnetic fields are no longer parallel to each other. 

Keeping terms only up to linear order in the electric field, and neglecting the terms involving gradient of $\Lambda_\mathbf{k}^\chi$ (as they are expected to be small), Eq.~\ref{Eq_boltz2} takes the following form 
\begin{align}
\mathcal{D}^\chi \left[v^{\chi z}_{\mathbf{k}} + \frac{e B}{\hbar} \sin \gamma (\boldsymbol{\Omega}^\chi\cdot \mathbf{v}^\chi_\mathbf{k})\right] = \sum\limits_{\eta}\sum\limits_{\mathbf{k}'} W^{\eta\chi}_{\mathbf{k}\mathbf{k}'} (\Lambda^{\eta}_{\mathbf{k}'} - \Lambda^\chi_\mathbf{k})
\end{align} 
In order to solve the above equation, we first define the scattering rate as follows
\begin{align}
\frac{1}{\tau^\chi_\mathbf{k}} = \mathcal{V} \sum\limits_{\eta} \int{\frac{d^3 \mathbf{k}'}{(2\pi)^3} (\mathcal{D}^\eta_{\mathbf{k}'})^{-1} W^{\eta\chi}_{\mathbf{k}\mathbf{k}'}}
\label{Eq_boltz3}
\end{align}
Substituting Eq.~\ref{Eq_W_1} in the above equation, we have 
\begin{widetext}
\begin{align}
\frac{1}{\tau^\chi_\mathbf{k}} = \frac{\mathcal{V}N}{8\pi^2 \hbar} \sum\limits_{\eta} |U^{\chi\eta}|^2 \iiint{(k')^2 \sin \theta' \mathcal{G}^{\chi\eta}(\theta,\phi,\theta',\phi') \delta(\epsilon^{\eta}_{\mathbf{k}'}-\epsilon_F)(\mathcal{D}^\eta_{\mathbf{k}'})^{-1}dk'd\theta'd\phi'},
\label{Eq_tau1}
\end{align}
\end{widetext}
where $N$ now indicates the total number of impurities, and $ \mathcal{G}^{\chi\eta}(\theta,\phi,\theta',\phi') = (1+\chi\eta(\cos\theta \cos\theta' + \sin\theta\sin\theta' \cos(\phi-\phi')))$ is the Weyl chirality factor defined by the overlap of the wavefunctions. The integration variable is changed from $k'$ to  $\epsilon^{\eta}_{\mathbf{k}'}$ by first defining the following constant energy contour $k^\eta$ in the momentum space 
\begin{align}
2\hbar v_F k^\eta &= \epsilon^\eta_{\mathbf{k}} +\sqrt{(\epsilon^\eta_{\mathbf{k}})^2 - 2 e \eta \hbar B v_F^2 \xi\kappa(\gamma,\theta,\phi)},
\end{align} 
where $\kappa(\gamma,\theta,\phi)=\cos\gamma \sin\theta \cos\phi + \sin\gamma \cos\theta$. Note that we have introduced another variable $\xi$ that can take the value 0 or 1. This enables us to keep a track of the explicit dependence on the orbital magnetic moment ($\xi=0$ kills all the OMM effects). The three-dimensional integral in Eq.~\ref{Eq_tau1} can then be reduced to just integration in $\phi'$ and $\theta'$. The scattering time ${\tau^\chi_\mathbf{k}}$ depends on  the Fermi energy or chemical potential ($\mu$), and is a function of variables $\theta$ and $\phi$. 
Specifically,
\begin{align}
\frac{1}{\tau^\chi_\mu(\theta,\phi)} = \mathcal{V} \sum\limits_{\eta} \beta^{\chi\eta}\iint{\frac{(k')^3}{|\mathbf{v}^\eta_{\mathbf{k}'}\cdot \mathbf{k}'^\eta|}\sin\theta'\mathcal{G}^{\chi\eta}(\mathcal{D}^\eta_{\mathbf{k}'})^{-1} d\theta'd\phi'},
\label{Eq_tau2}
\end{align}
where the prefactor $\beta^{\chi\eta} = N|U^{\chi\eta}|^2 / 4\pi^2 \hbar^2$, and the functional dependence of $\mathcal{G}^{\chi\eta}$ on the angular variables is implied but not explicitly mentioned for brevity. Finally, The Boltzmann equation (Eq~\ref{Eq_boltz3}) takes the following form  
\begin{align}
&h^\chi_\mu(\theta,\phi) + \frac{\Lambda^\chi_\mu(\theta,\phi)}{\tau^\chi_\mu(\theta,\phi)} =\nonumber\\ &\mathcal{V}\sum_\eta \beta^{\chi\eta}\iint {\frac{(k')^3}{|\mathbf{v}^\eta_{\mathbf{k}'}\cdot \mathbf{k}'^\eta|} \sin\theta'\mathcal{G}^{\chi\eta}(\mathcal{D}^\eta_{\mathbf{k}'})^{-1}\Lambda^\eta_{\mu}(\theta',\phi') d\theta'd\phi'}
\label{Eq_boltz4}
\end{align}
We will make the following ansatz for $\Lambda^\chi_\mu(\theta,\phi)$
\begin{align}
\Lambda^\chi_\mu(\theta,\phi) &= (\lambda^\chi - h^\chi_\mu(\theta,\phi) + a^\chi \cos\theta +\nonumber\\
&b^\chi \sin\theta\cos\phi + c^\chi \sin\theta\sin\phi)\tau^\chi_\mu(\theta,\phi)
\label{Eq_Lambda_1}
\end{align}
It is thus required to solve for eight unknowns ($\lambda^{\pm 1}, a^{\pm 1}, b^{\pm 1}, c^{\pm 1}$). The L.H.S in Eq.~\ref{Eq_boltz4} simplifies to $\lambda^\chi + a^\chi \cos\theta + b^\chi \sin\theta\cos\phi + c^\chi \sin\theta\sin\phi$. The R.H.S of Eq.~\ref{Eq_boltz4} assumes the form 
\begin{align}
\mathcal{V}\sum_\eta \beta^{\chi\eta} \iint &f^{\eta} (\theta',\phi') \mathcal{G}^{\chi\eta} (\lambda^\eta - h^\eta_\mu(\theta',\phi') + a^\eta \cos\theta' +\nonumber\\
	&b^\eta \sin\theta'\cos\phi' + c^\eta \sin\theta'\sin\phi')d\theta'd\phi',
	\label{Eq_boltz5_rhs}
\end{align}
where 
\begin{align}
f^{\eta} (\theta',\phi') = \frac{(k')^3}{|\mathbf{v}^\eta_{\mathbf{k}'}\cdot \mathbf{k}'^\eta|} \sin\theta' (\mathcal{D}^\eta_{\mathbf{k}'})^{-1} \tau^\chi_\mu(\theta',\phi')
\label{Eq_f_eta}
\end{align}
The above equations, when written down explicitly take the form of seven simultaneous equations to be solved for eight variables. The last constraint comes from the particle number conservation 
\begin{align}
\sum\limits_{\chi}\sum\limits_{\mathbf{k}} g^\chi_\mathbf{k} = 0
\label{Eq_sumgk}
\end{align}
Thus Eq.~\ref{Eq_Lambda_1}, Eq.~\ref{Eq_boltz5_rhs}, Eq.~\ref{Eq_f_eta} and Eq.~\ref{Eq_sumgk} can be solved together with Eq~\ref{Eq_tau2}, simultaneously for the eight unknowns ($\lambda^{\pm 1}, a^{\pm 1}, b^{\pm 1}, c^{\pm 1}$). Due to the complicated nature of the problem, the associated two dimensional integrals w.r.t \{$\theta'$, $\phi'$\}, and the solution of the simultaneous equations are all performed numerically. Note that the above procedure has to be repeated for each value of $\gamma$. 

Once the solution to the Boltzmann equation is obtained, we can then evaluate the linear response conductivities ($\sigma_{zz}$ and $\sigma_{xz}$). 
The current density is given by 
\begin{align}
\mathbf{j} = -\frac{e}{\mathcal{V}} \sum\limits_{\chi}\sum\limits_{\mathbf{k}} \dot{\mathbf{r}}^\chi f_\mathbf{k}^\chi
\end{align}
Substituting the expressions for $\dot{\mathbf{r}}^\chi$ and $f_\mathbf{k}^\chi$, we have
\begin{align}
\mathbf{j} = -e \sum\limits_{\chi} \frac{d^3 \mathbf{k}}{(2\pi)^3}& \left[\frac{e}{\hbar} (\mathbf{E} \times \boldsymbol{\Omega}^\chi) + \frac{e}{\hbar} (\mathbf{v}^\chi\cdot \boldsymbol{\Omega}^\chi) \mathbf{B} +\mathbf{v}^\chi\right]\nonumber \\ 
& \times	\left[f_0(\epsilon_\mathbf{k}^\chi)  - e \left(\frac{\partial f_0(\epsilon_\mathbf{k}^\chi)}{\partial \epsilon^\chi_\mathbf{k}}\right) E \Lambda^\chi\right]
\end{align}

From the above expression, the longitudinal conductivity ($\sigma_{zz}$) and the Hall conductivity ($\sigma_{xz}$) are evaluated (in the zero-temperature limit) to be 
\begin{align}
\sigma_{zz} &= -e \sum\limits_{\chi} \frac{1}{(2\pi)^2} \iint{\frac{(k^\chi)^3}{\hbar|\mathbf{v}^\chi_\mathbf{k}\cdot \mathbf{k}^\chi|}} \nonumber\\
&\times \left[-\frac{e^2}{\hbar} (\mathbf{v}^\chi\cdot \boldsymbol{\Omega}^\chi)  \Lambda^\chi B \sin \gamma - e \Lambda^\chi v^\chi_z \right]
\end{align}
\begin{align}
\sigma_{xz} &= -e \sum\limits_{\chi} \frac{1}{(2\pi)^2} \iint{\frac{(k^\chi)^3}{\hbar|\mathbf{v}^\chi_\mathbf{k}\cdot \mathbf{k}^\chi|}} \nonumber\\
&\times \left[-\frac{e^2}{\hbar} (\mathbf{v}^\chi\cdot \boldsymbol{\Omega}^\chi)  \Lambda^\chi B \cos \gamma - e \Lambda^\chi v^\chi_x \right]
\end{align}
We also define the coefficients ($\zeta_{xz}$ and $\zeta_{zz}$) that will be useful in discussing the behaviour of the above conductivities 
\begin{align}
\sigma_{xz} = \zeta_{xz} B^2 + \sigma^0_{xz} \label{Eq_sxz_zeta}\\ 
\sigma_{zz} = \zeta_{zz} B^2 + \sigma^0_{zz},
\label{Eq_szz_zeta}
\end{align}
where $\sigma^0_{zz}$ and $\sigma^0_{xz}$ are conductivities at zero magnetic field.
These coefficients are obtained numerically, solving the Boltzmann equation for each value of the ${B}-$field, and making a quadratic fit to the data. The linear-in-$B$ coefficients are expected to vanishe because the Weyl cones are untilted~\cite{sharma2017chiral}. 
Particularly, we will be interested in the sign of $\zeta_{zz}$ that directly relates to positive or negative LMC. Similarly the sign of $\zeta_{xz}$ will relate to increasing or decreasing planar Hall magnetoconductance (with respect to the applied magnetic field). 

\begin{figure*}
	\includegraphics[width=0.65\columnwidth]{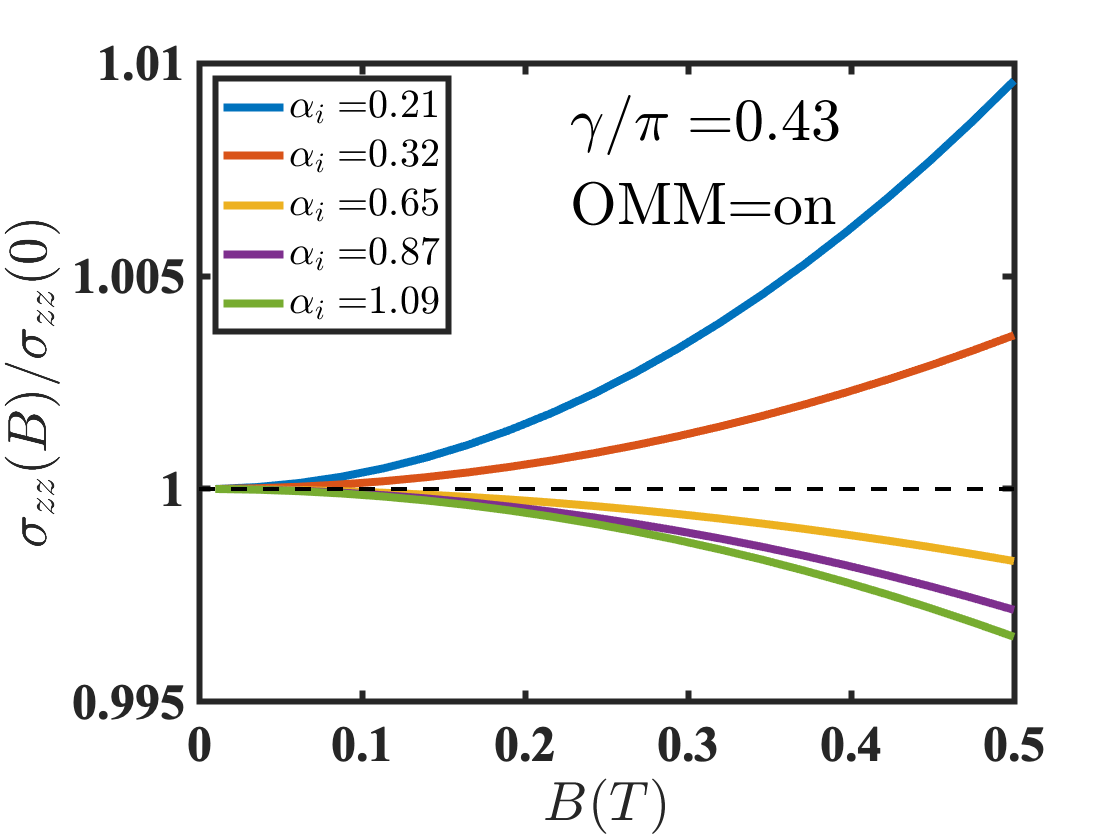}
	\includegraphics[width=0.65\columnwidth]{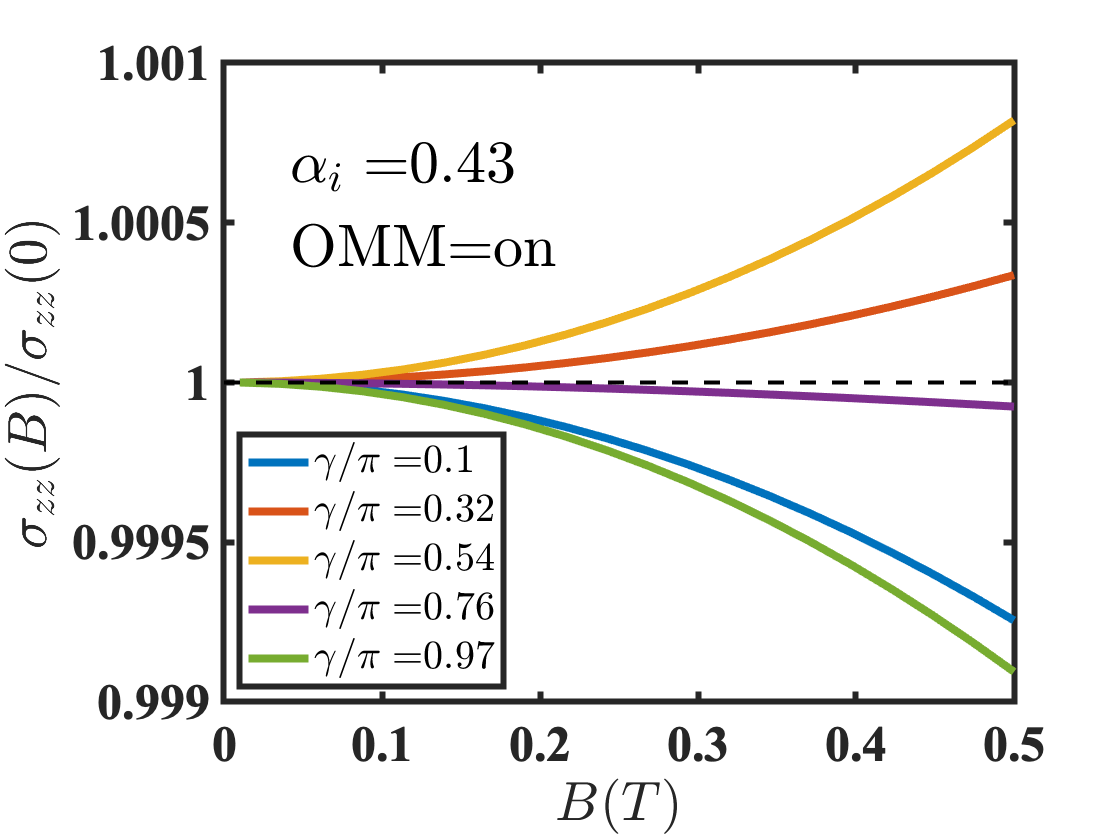}
	\includegraphics[width=0.65\columnwidth]{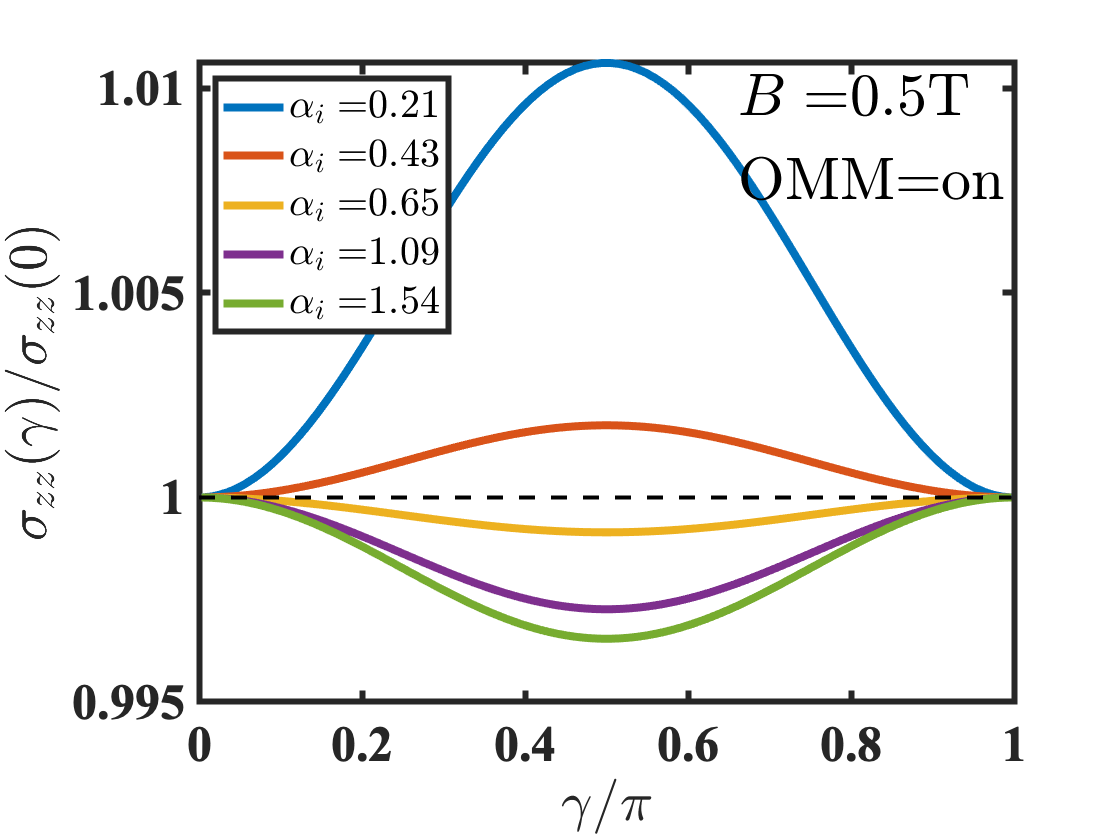}
	\caption{The behaviour of the longitudinal magnetoconductivity $\sigma_{zz}$ in the presence of orbital magnetic moment. \textit{Left:} $\sigma_{zz}$ vs. $B$ at a fixed direction of the magnetic field $\mathbf{B}$ (not parallel to $\mathbf{E}$) for various values of  the relative intervalley scattering $\alpha_i$. Beyond a critical $\alpha_i$ the chiral anomaly contribution to $\sigma_{zz}$ becomes negative as the LMC decreases with the magnetic field. \textit{Centre:} $\sigma_{zz}$ vs. $B$ at a fixed value of $\alpha_i$ for various directions of the magnetic field. As $\gamma$ varies from zero to $\pi/2$, we observe that the LMC contribution changes from negative to positive, and then vice-versa as $\gamma$ changes from $\pi/2$ to $\pi$. \textit{Right:} $\sigma_{zz}$ as a function of the direction of the magnetic field for various values of $\alpha_i$, roughly following the $\sin^2\gamma$ trend.  }
	\label{Fig1}
\end{figure*}

\begin{figure*}
	\includegraphics[width=0.65\columnwidth]{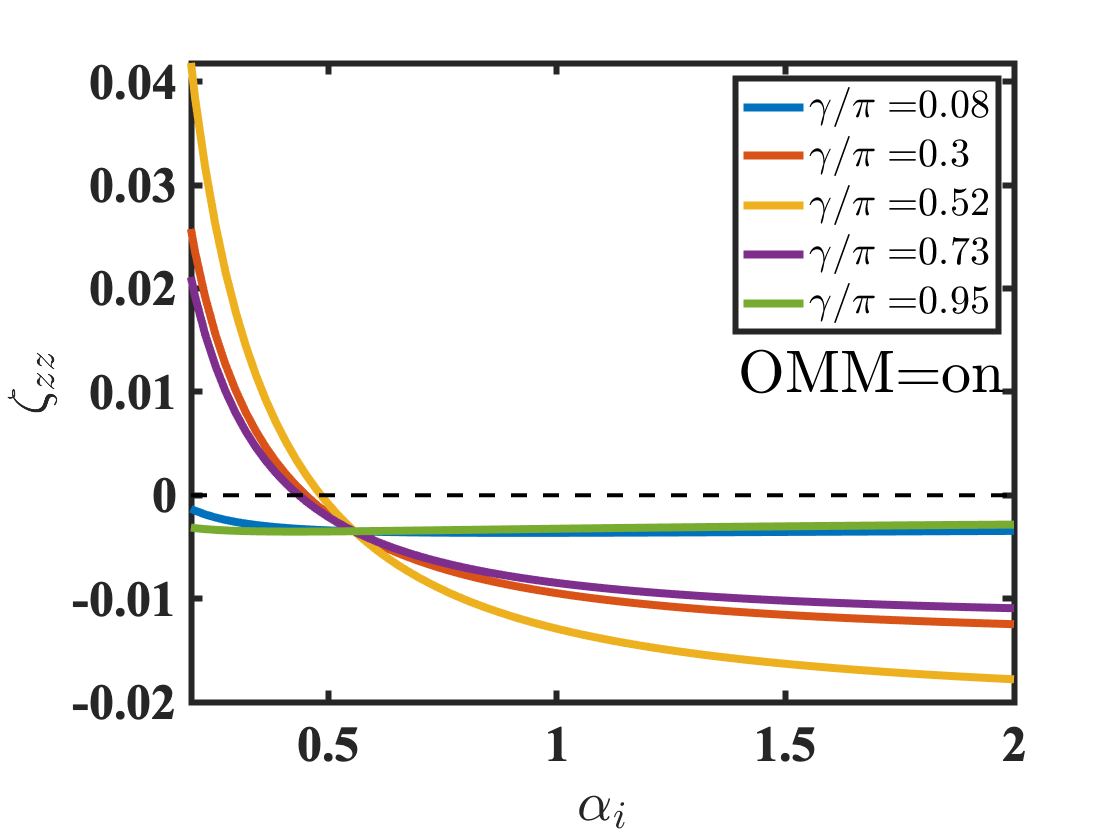}
	\includegraphics[width=0.65\columnwidth]{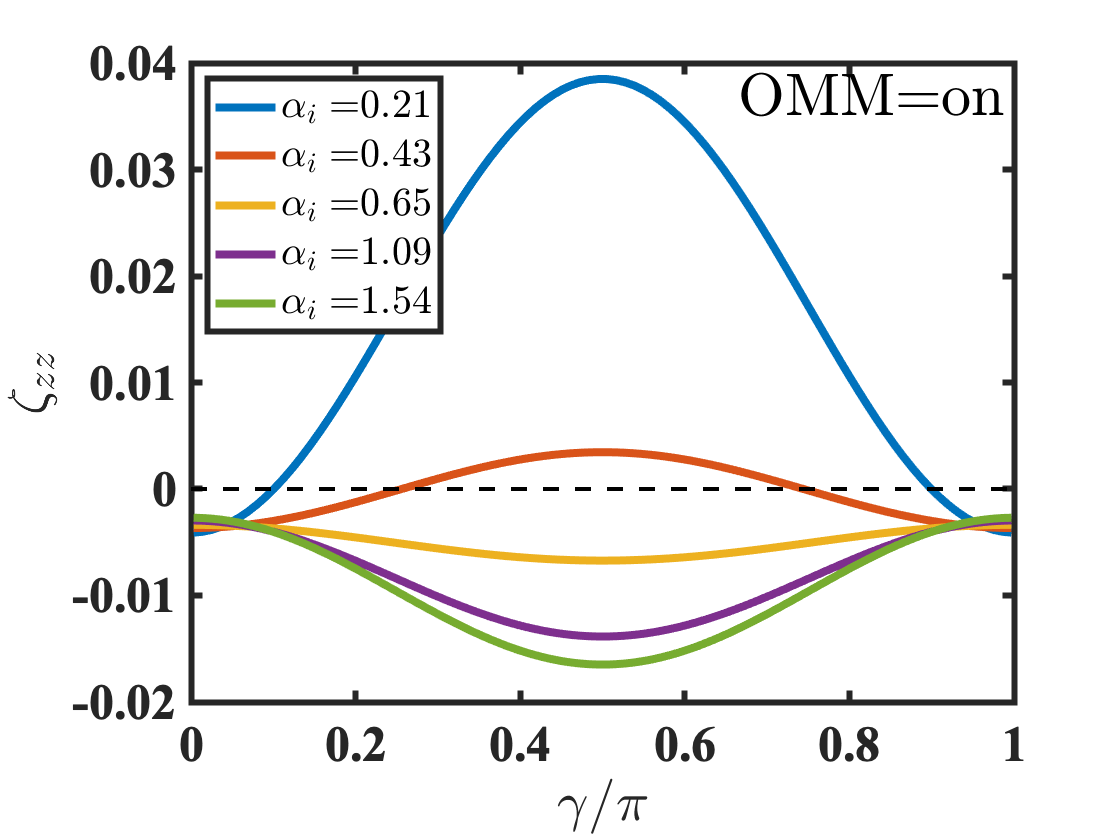}
	\includegraphics[width=0.65\columnwidth]{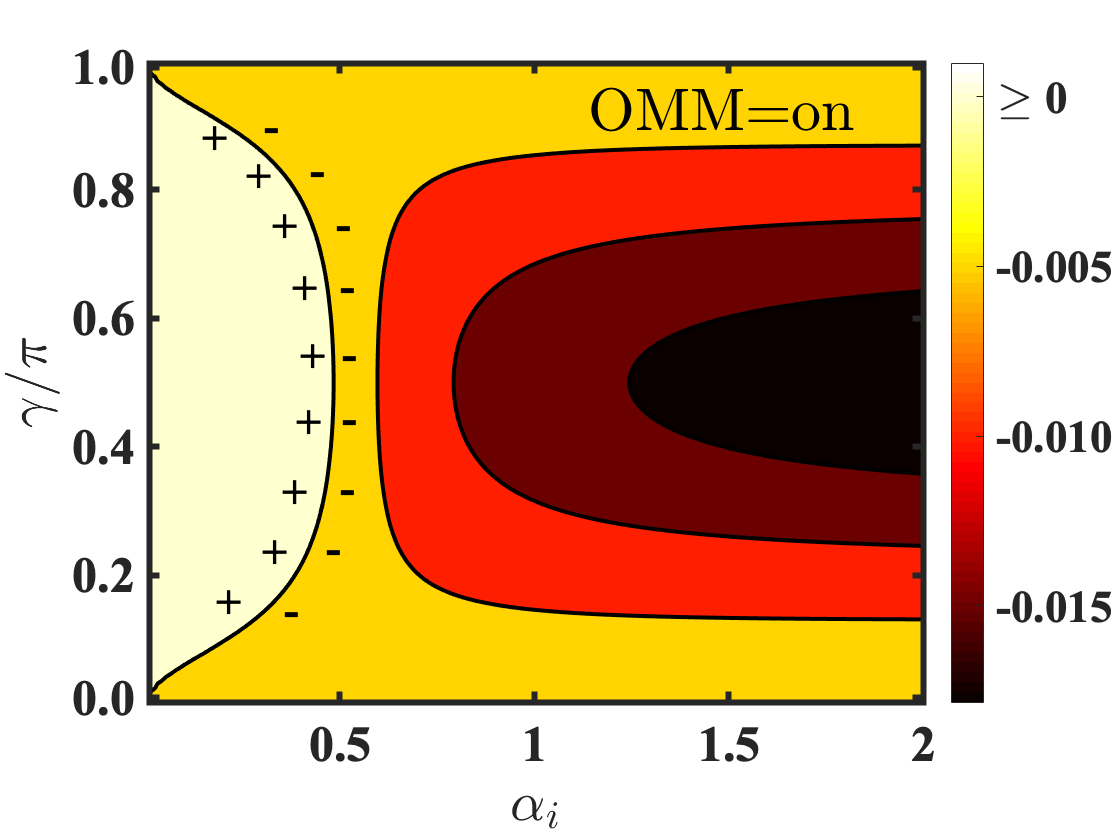}
	\caption{The coefficient $\zeta_{zz}$ (that explicitly indicates the sign of the LMC, see Eq.~\ref{Eq_sxz_zeta} and Eq.~\ref{Eq_szz_zeta}) in the presence of OMM. \textit{Left:} The coefficient $\zeta_{zz}$  as a function of $\alpha_i$ for various directions of the magnetic field. We note that $\zeta_{zz}$ changes sign at almost all directions of the magnetic field (except for $\gamma\rightarrow 0$ or $\gamma \rightarrow \pi$). The inversion feature is also well noted (i.e., the largest positive $\zeta_{zz}$ when $\alpha_i<\alpha_{ic}^\gamma$ becomes the largest negative when $\alpha_i>\alpha_{ic}^\gamma$ )  \textit{Centre:} The coefficient $\zeta_{zz}$ as a function of the direction of the $B-$field for various values of $\alpha_i$. For the intervalley scattering $\alpha_i$ below some critical strength, a sign change of the coefficient is observed as a function of $\gamma$. \textit{Right:} A contour plot of the the coefficient $\zeta_{zz}$ as a function of the direction of the applied $B-$field and the intervalley strength $\alpha_i$. We explicitly map the $\gamma-\alpha_i$ curve (the line separating $+/-$) in the parameter space, i.e., the $\alpha_{ic}^\gamma$ curve, where the LMC sign change occurs.}
	\label{Fig2}
\end{figure*}

\section{Results}
Having solved the Boltzmann equation, we now present the results for LMC as well as the planar Hall conductivity. In what follows we will denote the the ratio of intervalley scattering strength ($\beta^{\eta \eta'}$) to the intravalley scattering strength ($\beta^{\eta \eta}$) by $\alpha_i$, i.e., $\alpha_i = \beta^{+-}/\beta^{++}$. Further, valley symmetry is respected such that $\beta^{\eta \eta'} = \beta^{\eta' \eta}$. 
Fig.~\ref{Fig1} summarizes the behavior of LMC $\sigma_{zz}$ in the presence of OMM. First, at any particular angle $\gamma$,  we find that the LMC increases with the magnetic field until a critical value of $\alpha_i = \alpha_{ic}^{\gamma}$ is reached (the superscript $\gamma$ indicates that this critical value is angle dependent). For $\alpha_i>\alpha_{ic}^\gamma$, we find that the LMC decreases with increasing magnetic field. Second, for a fixed value of $\alpha_i$, the LMC can be either negative or positive depending on the angle $\gamma$. Further, we find that when $\gamma$ is closer to 0 or $\pi$, LMC is negative for even for infinitesimal intervalley scattering strength.  Lastly, when plotted as a function of angle, $\sigma_{zz}$ roughly is seen to follow the $\sin^2\gamma$ trend~\cite{nandy2017chiral} (at all values of $\alpha_i$).

\begin{figure*}
	\includegraphics[width=0.65\columnwidth]{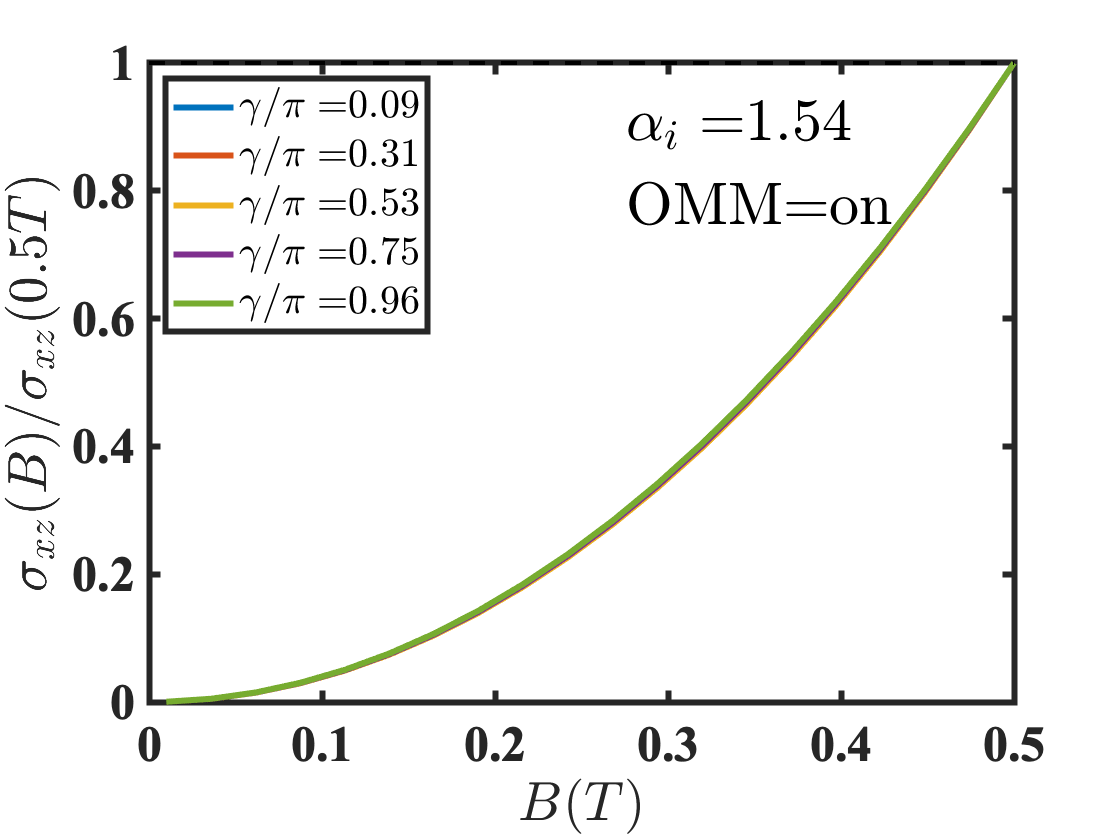}
	\includegraphics[width=0.65\columnwidth]{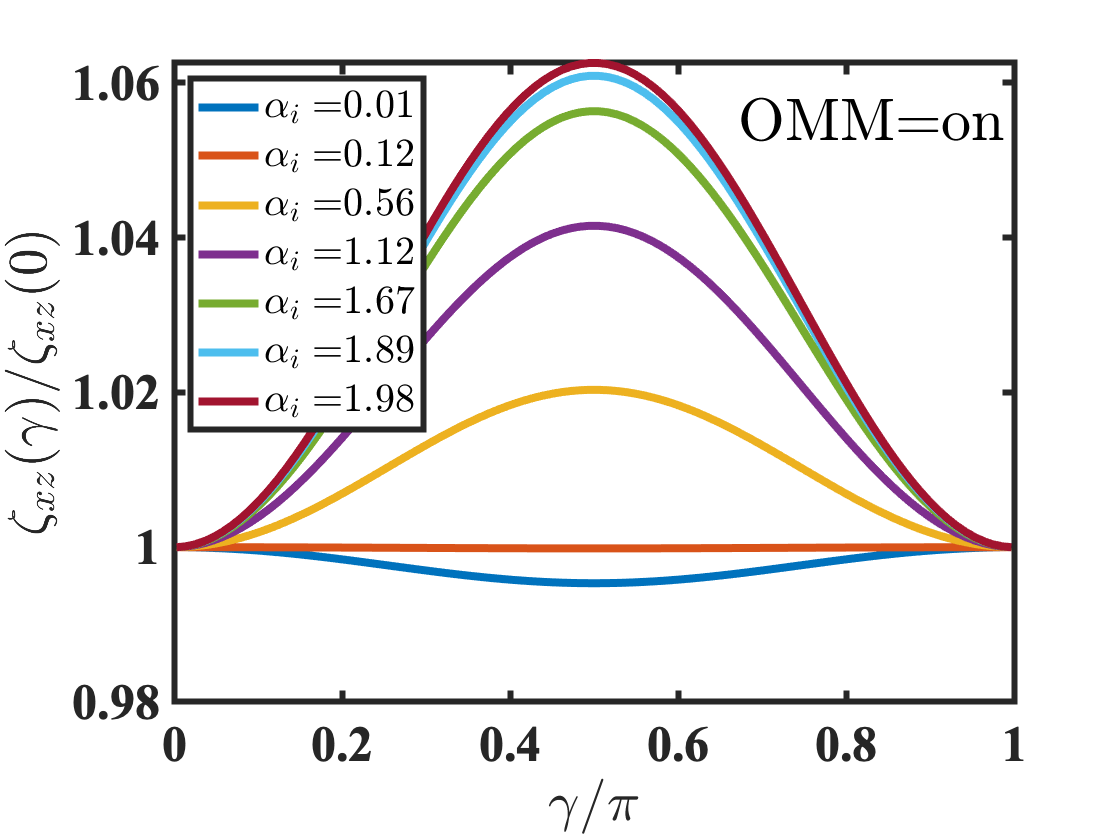}
	\includegraphics[width=0.65\columnwidth]{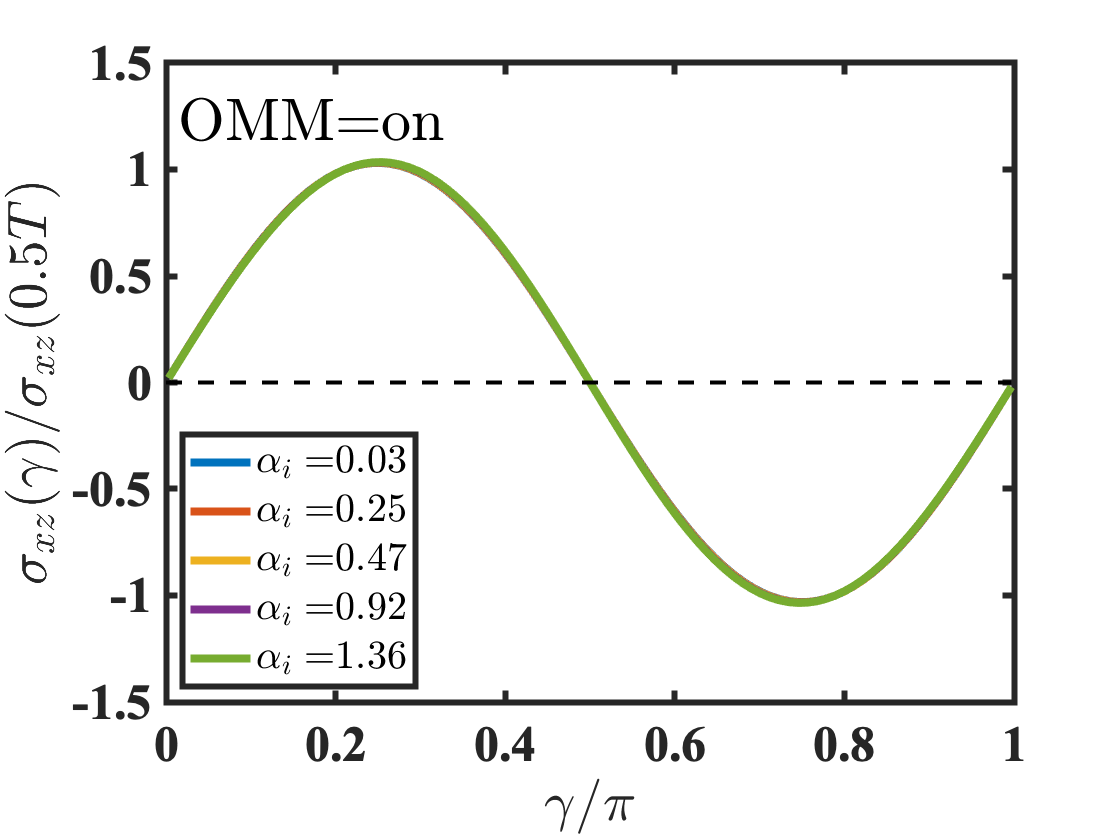}
	\caption{The behaviour of the planar Hall conductivity $\sigma_{xz}$ and the corresponding coefficient $\zeta_{xz}$ in the presence of an orbital magnetic moment. \textit{Left:} The planar Hall conductivity $\sigma_{xz}$ always increases with the applied magnetic field in $\sim B^2$ fashion even with strong intervalley scattering ($\alpha_i=1.54$) At each value of $\gamma$, the value of conductivity is normalized by it's value at $B=0.5T$. \textit{Centre:} The corresponding coefficient $\zeta_{xz}$ as a function of the direction $\gamma$ for different intervalley scattering strengths $\alpha_i$. No change of sign is observed in the coefficient at any value of $\gamma$ or $\alpha_i$ unlike LMC.  \textit{Right:} $\sigma_{xz}$ as a function of the magnetic field direction $\gamma$ following the $\sin\gamma\cos\gamma$ trend.  Here, the value of conductivity is normalized by it's value at $B=0.5T$ for each value of $\alpha_i$.}
	\label{Fig3}
\end{figure*}

\begin{figure*}
	\includegraphics[width=0.65\columnwidth]{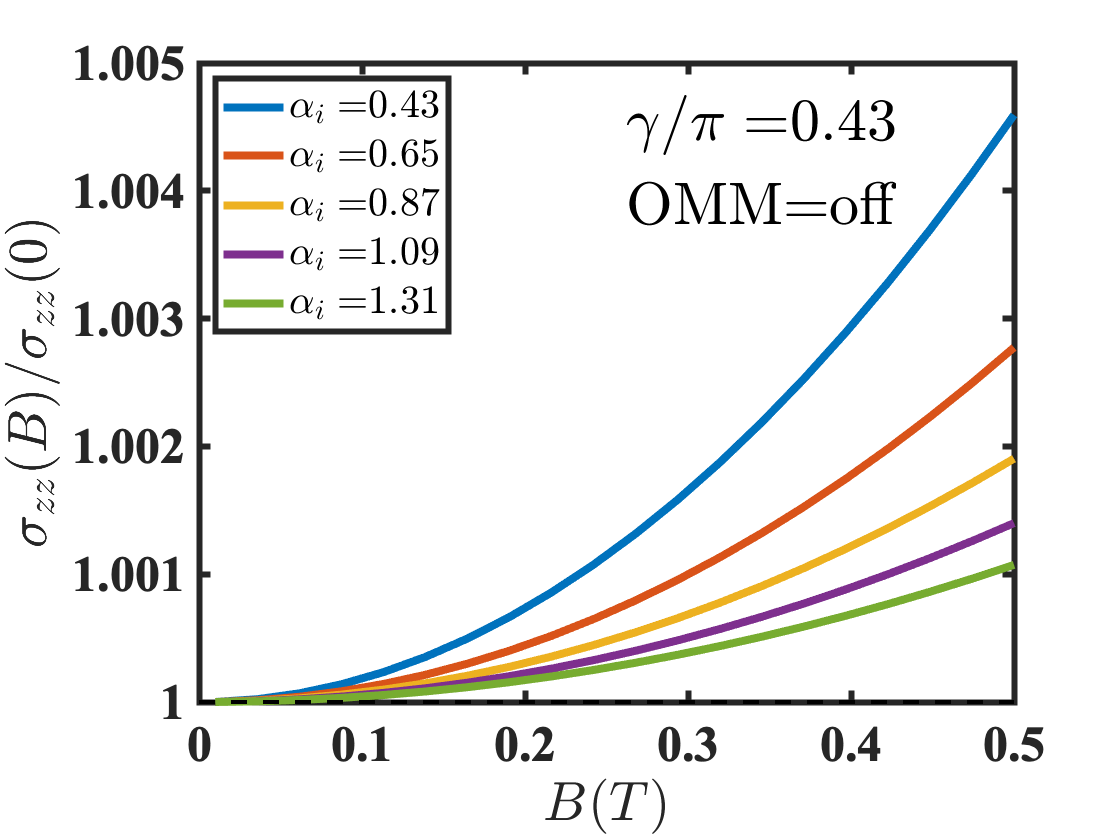}
	\includegraphics[width=0.65\columnwidth]{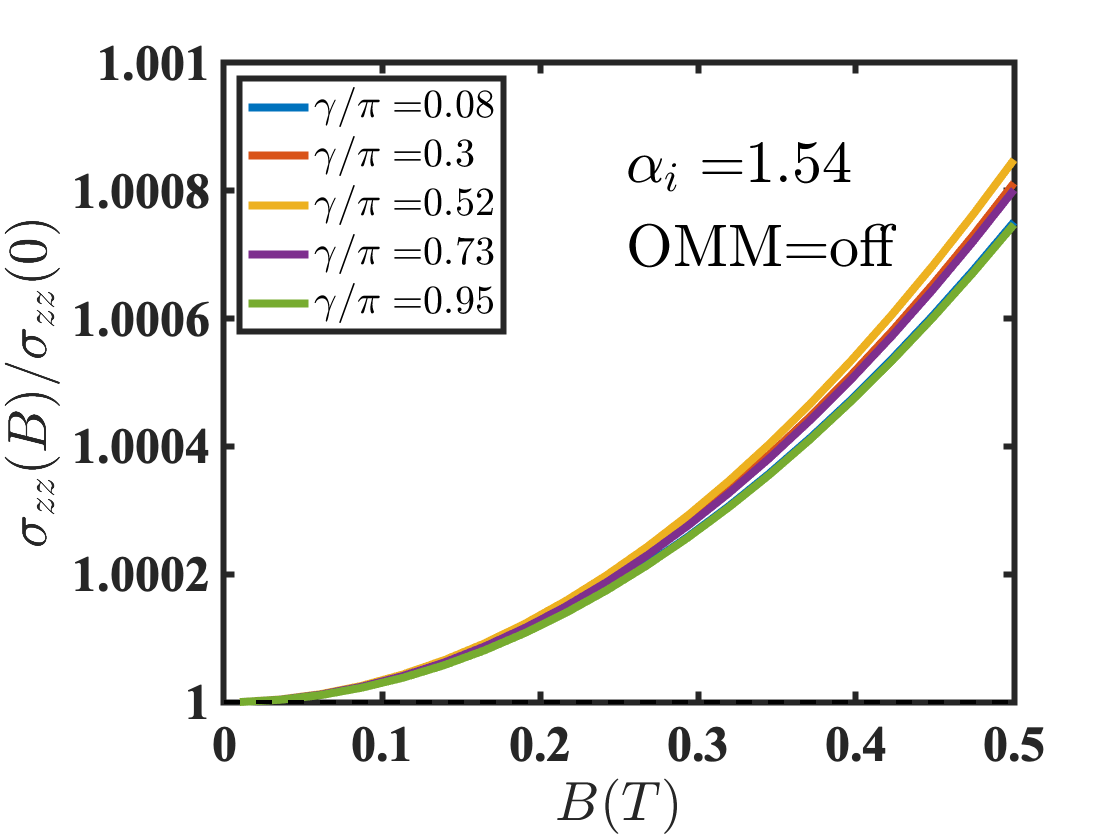}
	\includegraphics[width=0.65\columnwidth]{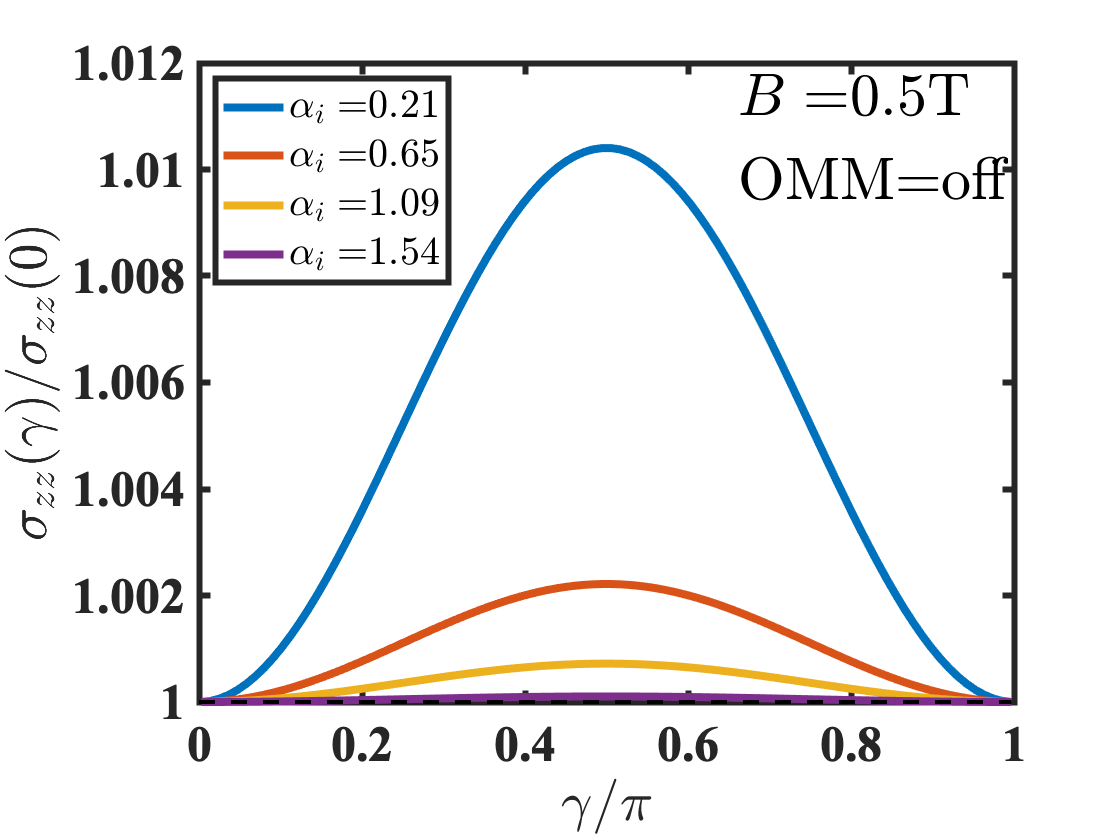}
	\caption{The behaviour of longitudinal magnetoconductivity $\sigma_{zz}$ in the absence of orbital magnetic moment. \textit{Left:} $\sigma_{zz}$ vs. $B$ at a fixed direction of the magnetic field $\mathbf{B}$ (not parallel to $\mathbf{E}$) for various values of  the relative intervalley scattering strength $\alpha_i$. The chiral anomaly contribution to $\sigma_{zz}$ is always positive. \textit{Centre:} $\sigma_{zz}$ vs. $B$ at a fixed value of (strong) $\alpha_i$ for various directions of the magnetic field. Again, the chiral anomaly contribution to $\sigma_{zz}$ is always positive. \textit{Right:} $\sigma_{zz}$ as a function of the direction of the magnetic field for various values of $\alpha_i$, roughly following the $\sin^2\gamma$ trend. }
	\label{Fig4}
\end{figure*} 

In order to get a more quantitative picture, we plot the coefficient $\zeta_{zz}$ in Fig.~\ref{Fig2}. As noted earlier, the sign of the coefficient $\zeta_{zz}$ directly relates to positive or negative LMC. The value $\alpha_{ic}^\gamma$ thus corresponds to the case when $\zeta_{zz}=0$. Starting with $\gamma$ close to 0, we find that $\zeta_{zz}$ is (i) small and (ii) negative for a wide range of $\alpha_i$. This is expected because if the electric and magnetic fields are almost perpendicular (though they need not be exactly orthogonal) to each other, the effects due to chiral anomaly are the least pronounced.  As $\gamma$ is slowly increased to $\pi/2$, $\zeta_{zz}$ increases with $\gamma$, and remains positive as long as $\alpha_i<\alpha_{ic}^\gamma$, again indicating that the chiral anomaly induced LMC are most pronounced when electric and magnetic fields are parallel to each other. When $\alpha_i$ crosses $\alpha_{ic}^\gamma$, we find a qualitative inversion of this feature, i.e.,  $\zeta_{zz}$ at $\gamma = \pi/2$ becomes the most negative when compared to other values of $\gamma$. This suggests that the manifestation of negative LMC is also a highlight of chiral anomaly, and not a suppression due to effects of OMM. Interestingly, we find that there exists a specific value of $\alpha_i>$ min $\{\alpha_{ic}^\gamma\}$, where the coefficient $\zeta_{zz}$ is the same for all the angles $\gamma$. Lastly, we also note that $\alpha_{ic}^\gamma$ is less sensitive to the actual value of $\gamma$ (as long as $\gamma$ is not too far away from $\pi/2$). These features are succinctly summarized in the contour plot (Fig.~\ref{Fig2} right) that maps the coefficient $\zeta_{zz}$ in the $\gamma-\alpha_i$ parameter space. We note that $\alpha_{ic}^\gamma$ is largely insensitive to $0.15\lesssim\gamma\lesssim0.85$, but then changes quickly for $0<\gamma\lesssim 0.15$ and $0.85\lesssim\gamma<\pi$.

Next, we discuss the planar Hall conductivity $\sigma_{xz}$ in the presence of orbital magnetic moment presented in Fig.~\ref{Fig3}. Even for large intervalley scattering ($\alpha_i=1.54$), the planar Hall conductivity $\sigma_{xz}$ is observed to always increase quadratically with the applied magnetic field irrespective of the angle $\gamma$. This is in contrast to LMC that was observed to decrease with the magnetic field at large values of $\alpha_i$ ($\alpha_i>\alpha_{ic}^\gamma$). As expected, the corresponding coefficient $\zeta_{xz}$ is always positive. Keeping the angle $\gamma$ fixed, as $\alpha_i$ is increased, $\zeta_{xz}$ is observed to monotonically increase with the scattering strength $\alpha_i$ and eventually saturates at large $\alpha_i$. As expected, the coefficient $\zeta_{xz}$ varies the most when $\mathbf{E}$ and $\mathbf{B}$ are parallel to each other. Finally, when plotted as a function of $\gamma$, $\sigma_{xz}$ resembles the $\sin\gamma \cos\gamma$ trend (Ref.~\cite{nandy2017chiral}), for all values of $\alpha_i$. 

\begin{figure*}
	\includegraphics[width=0.65\columnwidth]{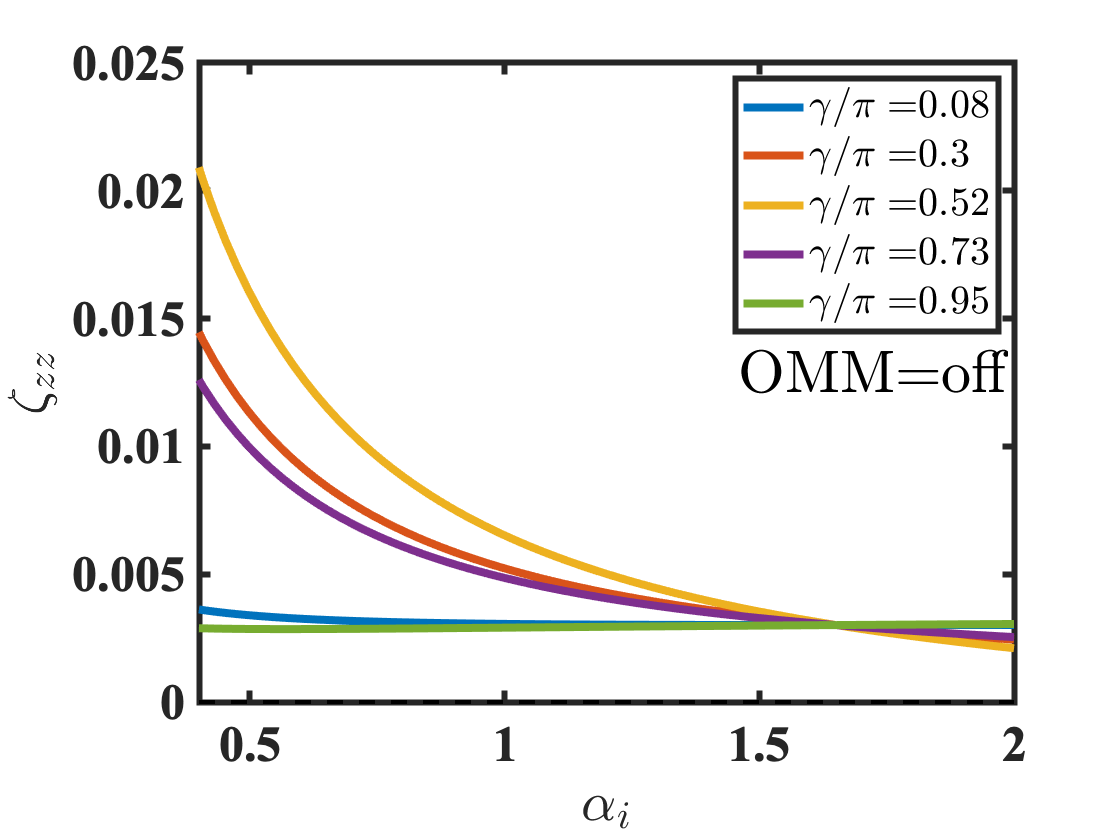}
	\includegraphics[width=0.65\columnwidth]{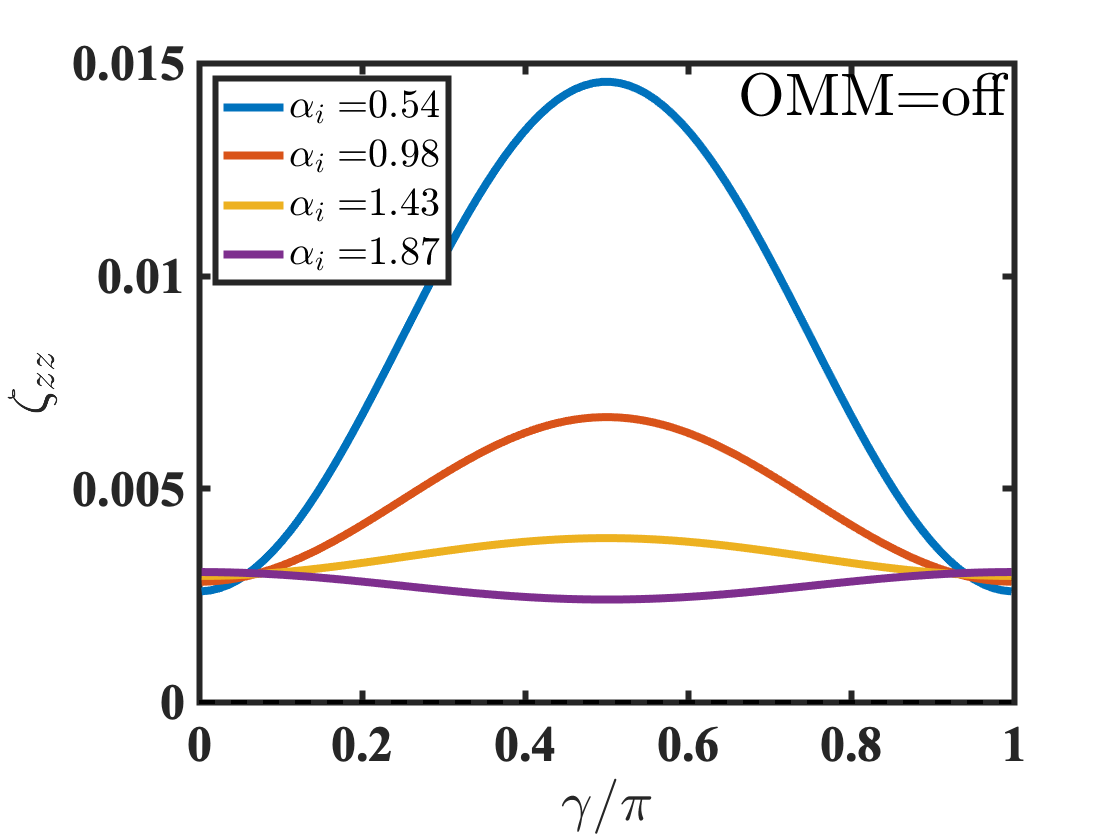}
	\includegraphics[width=0.65\columnwidth]{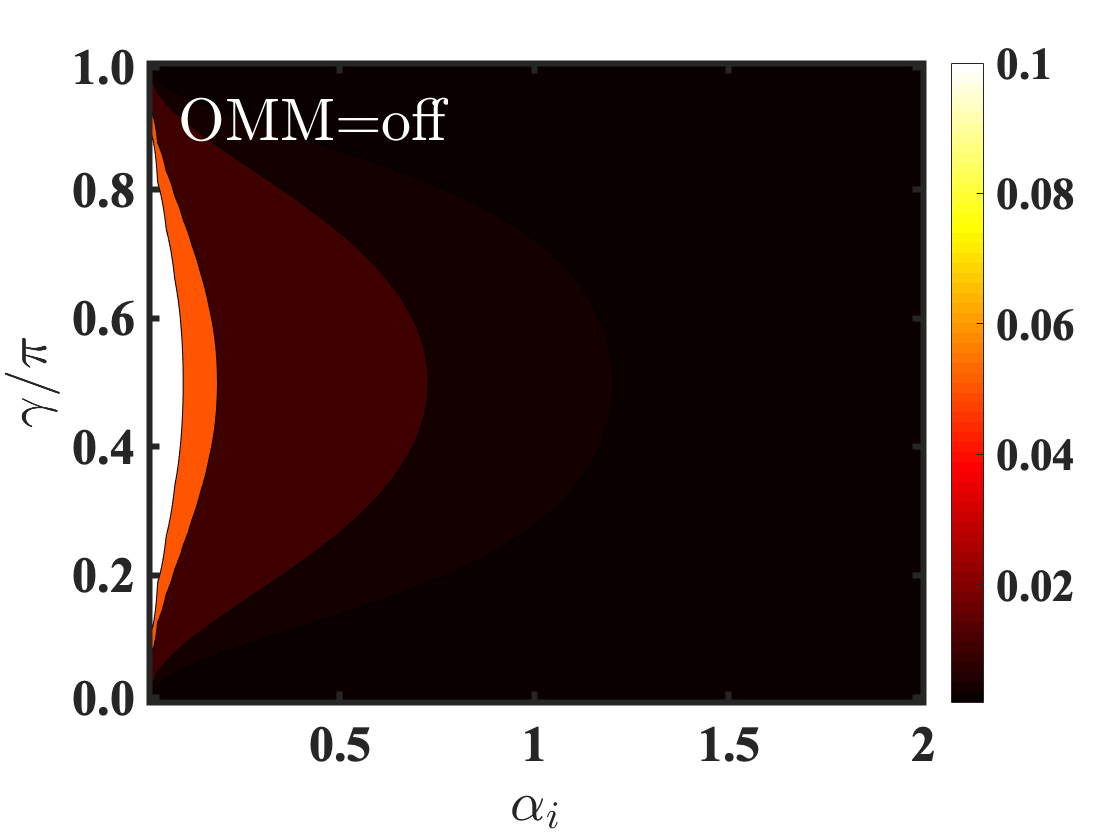}
	\caption{\textit{Left:} The coefficient $\zeta_{zz}$ (that explicitly indicates the sign of the LMC, see Eq.~\ref{Eq_sxz_zeta} and Eq.~\ref{Eq_szz_zeta}) as a function of $\alpha_i$ for various directions of the magnetic field. There is no change of sign an any value of $\alpha_i$. \textit{Centre:} The coefficient $\zeta_{zz}$ as a function of the direction of the $B-$field for various values of $\alpha_i$. Again, no change of sign is observed since the coefficient always remains positive. \textit{Right:} A contour plot of the the coefficient $\zeta_{zz}$ as a function of the direction of the applied $B-$field and the intervalley strength $\alpha_i$. Throughout the parameter space, $\zeta_{zz}$ is always positive, indicating that the sign of LMC is always positive. Thus the chiral anomaly contribution to $\sigma_{zz}$ is always positive in the absence of orbital magnetic moment.}
	\label{Fig5}
\end{figure*}

\begin{figure*}
	\includegraphics[width=0.65\columnwidth]{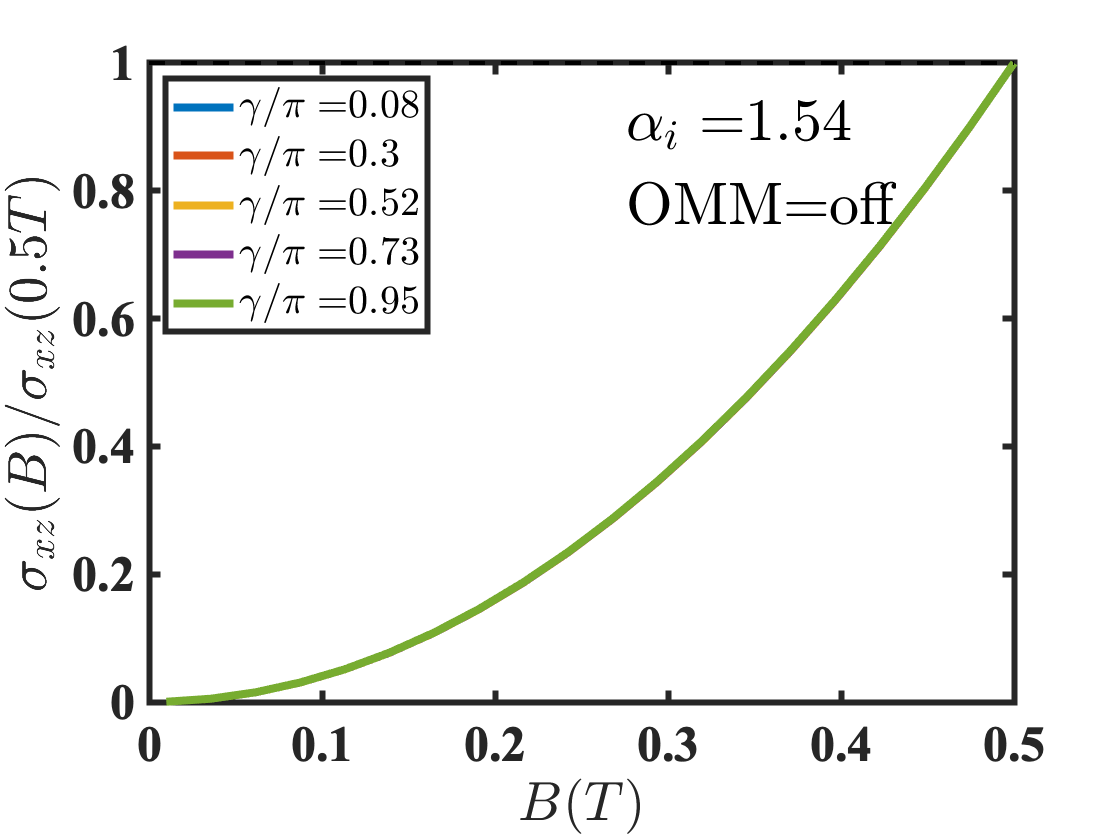}
	\includegraphics[width=0.65\columnwidth]{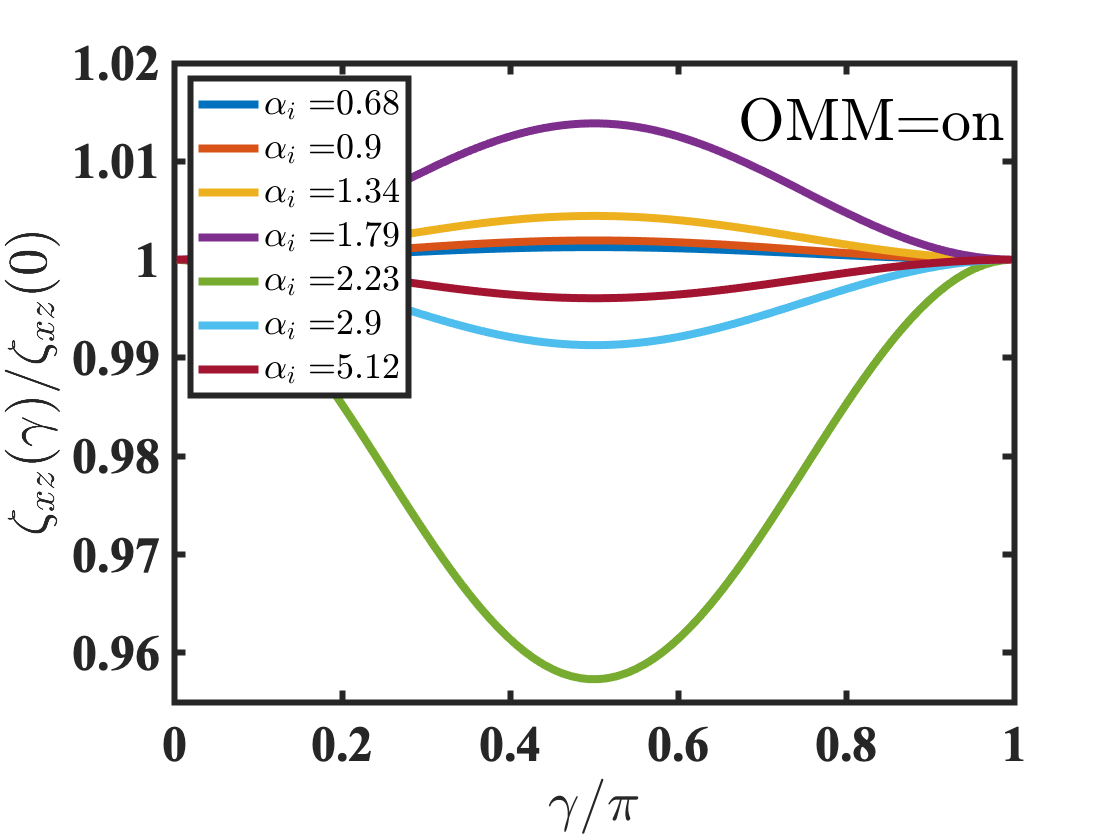}
	\includegraphics[width=0.65\columnwidth]{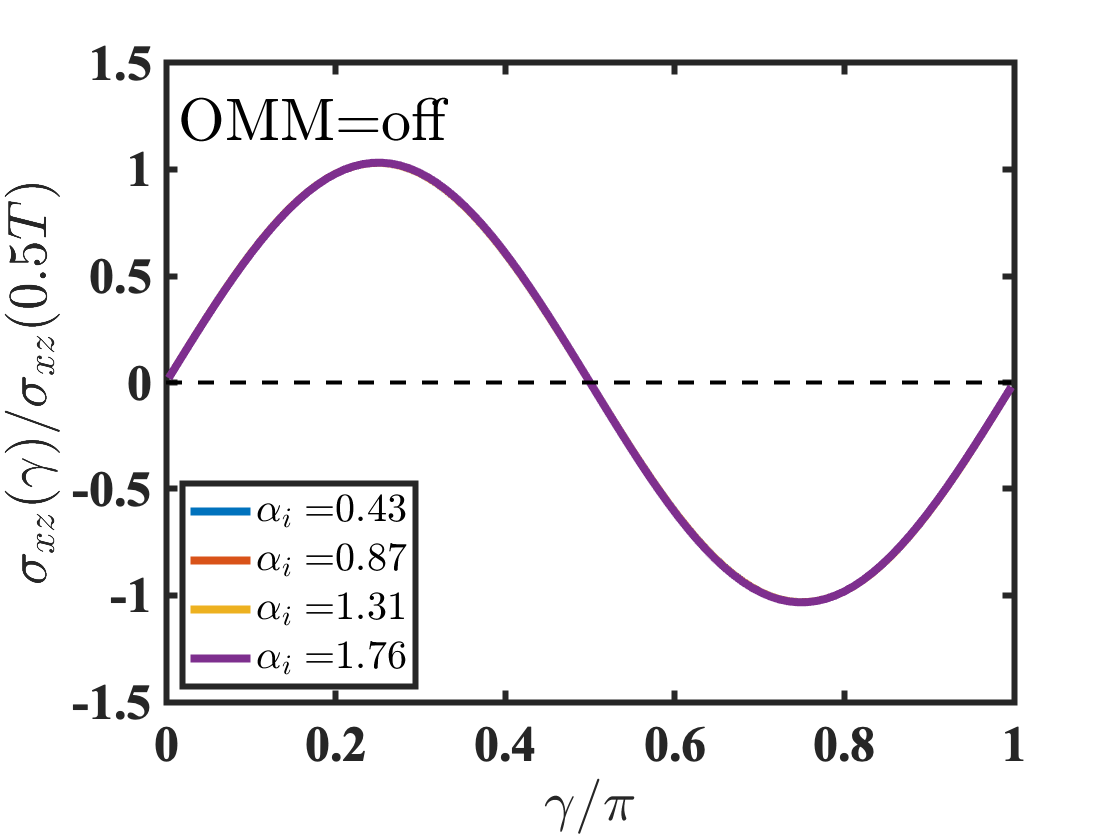}
	\caption{The behaviour of the planar Hall conductivity $\sigma_{xz}$ and the corresponding coefficient $\zeta_{xz}$ in the absence of orbital magnetic moment.  \textit{Left:} The planar Hall conductivity $\sigma_{xz}$ always increases with the applied magnetic field in $\sim B^2$ fashion even with strong intervalley scattering ($\alpha_i=1.54$) The value of conductivity is normalized by it's value at $B=0.5T$ for each value of $\gamma$. \textit{Centre:} The corresponding coefficient $\zeta_{xz}$ as a function of the direction $\gamma$ for different intervalley scattering strengths $\alpha_i$. The coefficient is always positive, indicating a positive contribution to PHE regardless of the direction of the magnetic field or the magnitude of the intervalley scattering strength. \textit{Right:} $\sigma_{xz}$ as a function of the magnetic field direction $\gamma$ following the $\sin\gamma\cos\gamma$ trend.  Here, the value of conductivity is normalized by it's value at $B=0.5T$ for each value of $\alpha_i$.}
	\label{Fig6}
\end{figure*}

We now move on to the discussion of LMC without including the effects of orbital magnetic moment, which is presented in Fig.~\ref{Fig4}. The first observation is that LMC is always observed to increase with respect to the external magnetic field, irrespective of the magnitude of $\alpha_i$, as well as the direction of the magnetic field $\gamma$ (both of which only have quantitative effects on the magnetoconductance). When plotted as a function of $\gamma$, the $\sin^2\gamma$ trend is observed like the previous case with OMM (Fig.~\ref{Fig1}). Now, if we compare the actual magnitudes, then we find that for $\alpha_i\ll$ min $\{\alpha_{ic}^\gamma\}$ (when LMC should be positive even with OMM), the LMC in the presence of OMM is larger than LMC in the absence of OMM. For example when $\alpha_i=0.43$, and $\gamma/\pi=0.43$, then $\sigma_{zz}(0.6T)/\sigma_{zz}(0)\sim 1.01$ with OMM, compared to $\sim 1.005$ without OMM. This suggests that effects associated with inclusion of the orbital magnetic moment work in conjunction with the intervalley scattering to enhance the signatures of chiral anomaly, and not suppress them.  

As a consequence, the coefficient $\zeta_{zz}$ always remains positive without OMM (Fig.~\ref{Fig5}.), although again, we do note that the actual magnitude of $\zeta_{zz}$ actually does decrease when OMM is absent (irrespective of the sign of $\zeta_{zz}$). Interestingly, even in this case we also find that there exists a particular value of $\alpha_i$, where the coefficient $\zeta_{zz}$ at all directions of the magnetic field become equal to each other. Beyond this conjunction point, we note the inversion feature (as also noted in Fig.~\ref{Fig2}), however, the inversion is far above the $\zeta_{zz}=0$ axis, and thus the coefficients always remain positive.
Finally, these features are summarized in the contour plot of $\zeta_{zz}$ in the $\gamma-\alpha_i$ parameter space. 

We finally discuss the results of the planar Hall conductivity in the absence of OMM as presented in Fig.~\ref{Fig6}. Even for large intervalley scattering ($\alpha_i=1.54$), the planar Hall conductivity $\sigma_{xz}$ is observed to always increase quadratically with the applied magnetic field (as the case with OMM in Fig.~\ref{Fig3}). As a consequence, the corresponding coefficient $\zeta_{xz}$ is always positive.  Keeping the angle $\gamma$ fixed, as $\alpha_i$ is increased, $\zeta_{xz}$ is initially seen to increase but then decreases after reaching a maximum (at around $\alpha_i \sim 1.8$). When $\alpha_i$ is increased further, $\zeta_{xz}$ again reaches a minimum (when $\alpha_i\sim 2.2$), and then increases again (the minimum never falls below zero). As expected, the coefficient $\zeta_{xz}$ varies the most when $\mathbf{E}$ and $\mathbf{B}$ are parallel to each other. 
Finally, when plotted as a function of $\gamma$, $\sigma_{xz}$ resembles the $\sin\gamma \cos\gamma$ trend, for all values of $\alpha_i$. 

\section{Discussion and Conclusions}
Unlike earlier claims in the literature, the observation of positive longitudinal magnetoconductance can no longer be thought of as a definitive signature of the manifestation of chiral anomaly in Weyl semimetals. As we show here (and as also shown recently~\cite{knoll2020negative}), chiral anomaly can result in both positive or negative LMC depending on the relative strengths of the intervalley and intravalley scattering. In fact, if only intervalley scattering is considered (i.e. $\alpha_i\gg 1$), we find that LMC must indeed always be {negative}, at least when the semiclassical conditions are met in the experiments. Thus concluding the manifestation of chiral anomaly, based only on the increase/decrease of LMC is by no means sufficient. 
Therefore, more definitive signatures are required to confirm whether the physical mechanism at play comes from chiral anomaly or something else. Some such signatures can arise when we consider a general configuration of the applied fields, i.e., the electric ($\mathbf{E}$) and magnetic ($\mathbf{B}$) fields and not necessarily parallel to each other. For example, as followed in this work, the direction of the electric field can be held fixed, and the $B-$field can be rotated, and the quantities $\sigma_{zz}$ (longitudinal conductivity) and $\sigma_{xz}$ (planar Hall conductivity) can be measured. 

To this end, we have solved the Boltzmann equation for a simple prototype of Weyl semimetal, in the presence of orbital magnetic moment, as well as both intravalley and intervalley scattering, in a general configuration when the applied fields are no longer parallel to each other (the angle between the fields related by the parameter $\gamma$, and $\gamma=\pi/2$ indicating both fields are parallel to each other, see Fig.~\ref{Fig_schematic}). We find that at a particular angle $\gamma$ (not too close to 0 or $\pi$),  the LMC increases and remains positive with the magnetic field until a critical value of the intervalley scattering $\alpha_i = \alpha_{ic}^{\gamma}$ is reached. For $\alpha_i>\alpha_{ic}^\gamma$, we find that the LMC decreases and becomes negative with increasing magnetic field. For a fixed value of $\alpha_i$, the LMC can be either negative or positive depending on the angle $\gamma$. Specifically, when $\gamma$ is close to 0 or $\pi$, LMC is typically always negative even for infinitesimal intervalley strength, but changes sign as $\gamma$ moves away from 0 or $\pi$. These features are also discussed in terms of a related quantity (the coefficient $\zeta_{zz})$, that gives us the estimate of the magnitude and sign of chiral anomaly contribution. 
When $\alpha_i$ crosses $\alpha_{ic}^\gamma$, we find a qualitative inversion of $\zeta_{zz}$, i.e., $\zeta_{zz}$ is the largest positive at $\gamma = \pi/2$ when $\alpha_i<\alpha_{ic}$, and becomes the largest negative when $\alpha_i>\alpha_{ic}$, compared to other values of $\gamma$. This suggests that the manifestation of negative LMC is also a highlight of chiral anomaly (just like positive LMC), because chiral anomaly effects are pronounced when the $\mathbf{E}$ and $\mathbf{B}$ fields are parallel to each other. 

This angular dependence of the sign of LMC (change from positive to negative) can be regarded as a key signature of the underlying specific mechanism at play (namely, chiral anomaly). This is expected to be true because in a generic metal without Weyl nodes, LMC is generally expected to remain negative and not change sign as the direction of the applied $B-$field is rotated. On the other hand, even if positive LMC is observed without Weyl nodes (as in PdCoO$_2$), again, a change of sign is unlikely, if the $B-$field is rotated. We have also explicitly mapped out the phase diagram of the coefficient $\zeta_{zz}$ in the $\alpha_i-\gamma$ parameter space, which  traces out the curve $\alpha_{ic}^\gamma$, that can also be mapped experimentally. 
Additionally, we also explore the planar Hall conductivity ($\sigma_{xz}$) in this setup. We find that $\sigma_{xz}$, unlike LMC, always increases with a positive chiral anomaly contribution, as a function of the applied magnetic field, irrespective of the direction of the magnetic field and the strength of the relative intervalley strength $\alpha_i$ (showing only quantitative variations with respect to both these parameters). Finally, for completeness, we solve for both $\sigma_{zz}$ and $\sigma_{xz}$, excluding the effects of the orbital magnetic moment, and find that the chiral anomaly contribution always increases with the magnetic field irrespective of the angle $\gamma$ and the intervalley scattering strength $\alpha_i$. Our predictions can be directly tested in experiments, and may be employed as new diagnostic procedures to verify chiral anomaly in Weyl semimetals.

\textit{Acknowledgements:} ST acknowledge support from ARO Grant No. W911NF16-1-0182. GS acknowledges IIT Mandi startup funds.
\bibliography{biblio}
\end{document}